\documentclass[preprint,5p,times,twocolumn]{elsarticle}

\usepackage{comment}
\usepackage{graphicx}               
\usepackage{paralist}        
\usepackage{tabularx}           
\usepackage{comment}
\usepackage{balance}          
\usepackage{multirow}      
\usepackage{fancyvrb}
\usepackage[TABBOTCAP]{subfigure} 
\usepackage{balance}
\usepackage{amsmath}
\usepackage{listings} 
\usepackage{makecell}
\usepackage{booktabs} 
\usepackage[table]{xcolor}
\usepackage{lscape}
\usepackage{xspace}
\usepackage{color, colortbl}
\usepackage[hyphens]{url}
\usepackage{hyperref}
\hypersetup{breaklinks=true}

\def\BibTeX{{\rm B\kern-.05em{\sc i\kern-.025em b}\kern-.08em
    T\kern-.1667em\lower.7ex\hbox{E}\kern-.125emX}}
    
\newcommand{\ie}{\textit{i.e.,}\xspace}
\newcommand{\eg}{\textit{e.g.,}\xspace}

\newcommand{\etal}{\textit{et al.}\xspace}
\newcommand{\revised}[1]{{\color{black}{#1}}}
\newcommand{\revise}[1]{{\color{black}{#1}}}
\newcommand{\change}[1]{{\color{black}{#1}}}
\newcommand{\major}[1]{{\color{black}{#1}}}

\usepackage{paralist}
\usepackage{paralist}
\usepackage{color}
\usepackage{url}

\makeatletter
\def\ps@pprintTitle{%
  \let\@oddhead\@empty
  \let\@evenhead\@empty
  \let\@oddfoot\@empty
  \let\@evenfoot\@oddfoot
}
\makeatother

\begin{document}

\title{Profiling Gas Consumption in Solidity Smart Contracts}

\author[RCOST]{Andrea Di Sorbo}
\ead{disorbo@unisannio.it}
\author[RCOST]{Sonia Laudanna}
\ead{slaudanna@unisannio.it}
\author[RCOST]{Anna Vacca}
\ead{avacca@unisannio.it}
\author[RCOST]{Corrado A. Visaggio}
\ead{visaggio@unisannio.it}
\author[RCOST]{Gerardo Canfora}
\ead{canfora@unisannio.it}
\address[RCOST]{Department of Engineering, University of Sannio, Italy}






\begin{abstract}
Nowadays, more and more applications are developed for running on a distributed ledger technology, namely dApps. The business logic of dApps is usually implemented within smart contracts developed through Solidity, a programming language for writing smart contracts on different blockchain platforms, including the popular Ethereum. In Ethereum, the smart contracts run on the machines of miners and the gas corresponds to the execution fee compensating such computing resources. However, the deployment and execution costs of a smart contract depend on the implementation choices done by developers. Unappropriated design choices could lead to higher gas consumption than necessary. In this paper, we (i) identify a set of 19 Solidity code smells affecting the deployment and transaction costs of a smart contract, and (ii) assess the relevance of such smells through a survey involving 34 participants. On top of these smells, we propose GasMet, a suite of metrics for statically evaluating the code quality of a smart contract from the gas consumption perspective. An experiment involving 2,186 smart contracts demonstrates that the proposed metrics have direct associations with deployment costs. The metrics in our suite can be used for more easily identifying source code segments that need optimizations. 
\end{abstract}


\maketitle

\section{Introduction}
\label{sec:intro}
Blockchain emerged as the enabling technology~\cite{satoshi} for implementing transactions of electronic cash, namely \textit{cryptocurrency}, without the brokerage of a financial institution. Thanks to the versatility of this technology, it is now increasingly adopted in several contexts, with different purposes far beyond crypto-currencies.
Indeed, blockchain is impacting a variety of business sectors; at illustrative aim, some recent applications using blockchain concern\footnote{See \url{https://builtin.com/blockchain/blockchain-applications}}: sharing of sensitive data, electronic voting, cross-border payments, personal identity security, goods authenticity and traceability, real estate processing. 
The global blockchain technology market size is expected to reach USD 57,641.3 million by 2025~\cite{blockchainreport}. 
This forecast entails that in the next years we will witness a significant spreading of decentralized applications \major{(dApps)}
, \textit{i.e.,} computer applications which run on a distributed ledger technology (DLT).

The business logic of dApps is usually implemented within smart contracts, \textit{i.e.,} fully-fledged programs that run on blockchain. Ethereum is one of the most popular blockchain platforms~\cite{chen2017under} and provides an environment to code and run smart contracts~\cite{DBLP:conf/wcre/WohrerZ18}. In this environment, smart contracts are typically developed through the Solidity object-oriented language, before being compiled into bytecodes that can be executed by the Ethereum Virtual Machine (EVM). 

As each underlying technology imposes its peculiarities and characteristics to applications that run on top of it, also blockchain introduces critical aspects affecting smart contract development~\cite{DBLP:journals/jss/VaccaSVC21}. 
For instance, the immutability of blockchain complicates smart contract maintenance activities, as, once deployed, they can hardly be patched~\cite{destefanis2018smart}. For this reason, it is crucial ensuring that a smart contract is bug-free and well designed \textit{before} deploying it to the blockchain~\cite{chen2020defining}. Besides, since smart contracts run on a blockchain infrastructure, a key factor is the cost of execution due to the mining of the blocks participating in the chain. On the Ethereum platform, the gas (in Ether) corresponds to the execution fee compensating for such computing resources. Pragmatically speaking, gas corresponds to real money; unoptimized smart contracts can lead to unnecessary gas leaks and, thus, to money losses~\cite{marchesi2020design}. Moreover, when the users produce a high number of transactions, if the dApp is not properly designed, the execution costs could be not sustainable by the business model of the Decentralized Application Organization (DAO) which manages it. 

The choices done by developers, such as the types of data structures used, the number of cycles, the kind of instructions, the types of variables used, where and how they are initialized or valued, may affect the gas consumption of a smart contract. Although recent research~\cite{marchesi2020design} outlined design patterns and guidelines for the development of more optimized smart contracts from the gas consumption perspective, programming resources and development environments that can help developers more easily identifying code units that need to be optimized are still lacking.       
This is also confirmed by the survey conducted by Zou et al.~\cite{zou2019smart}, in which the majority of interviewed smart contract developers feel that optimizing gas is painful, especially in complex applications, and it would be important to have tools that allow optimizing smart contract source code rather than bytecode. Indeed, to estimate gas consumption, a typical workflow consists of deploying the smart contract within a simulated DLT running on \major{a} private  workstation or local servers and obtaining an estimation of the cost\footnote{\url{https://ethereum.stackexchange.com/questions/27452/how-to-estimate-gas-cost}}. If the cost is too high, the smart contract is modified, deployed again on the local simulated DLT, and the new cost is evaluated. This process is repeated till the result is satisfying. 
\change{Recently, Ajienka et al.~\cite{DBLP:journals/smr/AjienkaVC20} studied the correlation between object-oriented metrics and the resources required for smart contracts deployment. The results of this prior work demonstrate that while inheritance-based metrics represent good indicators of the gas used for smart contracts deployment, coupling-related metrics do not. This achievement partially confirms the belief that traditional software metrics do not capture all the specific aspects of smart contracts and the need of metrics specifically designed for smart contracts~\cite{ducasse2019open}. }

To fill this gap, the goal of our work is to provide a tool for helping developers more easily identifying code units that may be optimized for achieving gas savings. \major{In particular, developer communities have identified Solidity \textit{cost smells}, which are coding practices that can negatively affect deployment and transaction costs of a smart contract.} Such cost smells are not gathered into a unique catalog, but they are dispersed within different books, reports, and web forums. In this paper, we collect a set of 19 cost smells and, through a survey involving 34 among smart contract developers and researchers, assess the perceived relevance of the collected cost smells. \change{The surveyed developers discussed the extent to which they agree that the identified smells are actual problems and that the proposed solutions are potential candidates to fix those problems, achieving an average agreement of about 70\% over the smells and solutions presented in the catalog.}
On top of these smells, we define a suite of code metrics, namely \textsc{GasMet} (standing for \textit{Gas Metrics}), for statically evaluating the code quality of smart contracts, from the gas consumption perspective. 
Through an experiment involving 2,186 real-world smart contracts, we obtain empirical evidence about the relationships existing between the defined metrics and gas consumption required by the deployment of a smart contract. 
The advantage brought by GasMet is that the developers can \revise{more easily} localize the segments of their code that need optimization, \textit{while} coding, without \revise{deploying or running} it on a DLT, simulated or real. Thus, the metrics in our suite can help developers implementing more optimized smart contracts requiring lower resource consumption. \major{In particular, the collection of GasMet metrics can give developers general indications of gas consumption inefficiencies occurring in the code, fostering the application of gas-saving patterns~\cite{marchesi2020design}.}
Besides, blockchain researchers could leverage the proposed metrics  to develop linters for accurately capturing the occurrences of cost smells in source code. 

The original contribution of this paper includes:
\begin{itemize}
    \item \emph{\major{a collection of cost smells, \textit{i.e.,} coding practices that entail a higher cost of smart contract deployment and execution;}}
    \item \emph{the results of a survey involving real smart contract developers on the perceived relevance of each defined cost smell;}
    \item \emph{GasMet, a suite of metrics for identifying the occurrences of the defined cost smells; and} 
    \item \emph{an empirical study for identifying the GasMet metrics that most correlate with deployment costs}.
\end{itemize}


\major{Previous research proposed tools based on symbolic execution for (i) automatically inferring gas upper bounds of smart contracts' public functions to prevent out-of-gas vulnerabilities~\cite{GASTAP}, or (ii) analyzing the number and types of bytecode instructions executed to detect under-optimized storage patterns~\cite{GASOL}, whereas static analysis techniques turned out to be effective for automatically identifying gas-related vulnerabilities~\cite{MadMax}. Our suite of metrics (i) provides information that is complementary to the one provided by the aforementioned tools, and (ii) \major{aims at supporting developers in identifying potential inefficiencies} in the smart contracts/functions implementations that could lead to higher gas consumption when deploying smart contracts. In particular, our metrics are related to a variety of gas-inefficient practices (not only to storage usage) and they are computed directly on the Solidity source code. In addition, our metrics do not take into account function inputs, as previous experiments demonstrated that only a small percentage of smart contracts (\ie about 10\%) do not follow the Ethereum safety recommendations\footnote{https://github.com/ethereum/wiki/wiki/Safety} and their gas consumption depends on the size of data stored, the size of functions inputs, or the blockchain state~\cite{GASTAP}.}

\textbf{Paper structure.} The paper proceeds as follows. Section~\ref{sec:Related} highlights the novelty of our findings with respect to the existing literature. Section~\ref{sec:elicitation} details the identified \textit{cost smells}, while Section~\ref{sec:studyRelevance} illustrates our study on the relevance of the identified cost smells and discusses the related results. Section~\ref{sec:suite} introduces the GasMet suite \major{and tool}, while Section~\ref{sec:design} deals with the \change{evaluation} of the suite and the results of our correlation analysis with gas consumption. Threats to our study's validity are commented on in Section~\ref{sec:threats} and, finally, Section~\ref{sec:conclusions} concludes the paper. 

\section{Related work}
\label{sec:Related}
In this section, we discuss the related literature that has been mainly devoted to studying (i) software metrics for Blockchain-Oriented Software (BOS in the remainder of the paper), (ii)  issues related to weaknesses in smart contracts source code, and (iii) evaluation of gas consumption.

\subsection{BOS software metrics}
Ortu \etal~\cite{ortu2019comparing} provided a statistical characterization of BOS. The authors inspected and compared 5 C++ open source Blockchain-Oriented and 5 Traditional Java software systems, to discover strength differences between the two categories of projects, and particularly in the statistical distribution of 10 software metrics. Even if there are similarities between the statistical distributions for Traditional software and Blockchain software, the distribution of Average Cyclomatic and Ration Comment To Code metrics detect significant differences, whereas the Number of Statements metric reveals meaningful differences on the double Pareto distribution. 

Hegedus~\cite{hegedHus2019towards} used \major{Object-Oriented (OO)} metrics for estimating properties of the smart contracts written in Solidity. Based on the results, the author found that smart contract programs are short, \major{not excessively complex and with few or no comments.} Furthermore, it would be useful for smart contracts to have an external library and dependency management mechanisms because many libraries have similar functionalities. 

Tonelli \etal~\cite{tonelli2018smart}
investigated the differences between the software metrics measured on Smart contracts (SC) and metrics extracted from traditional software systems. \major{The authors built their dataset from the Etherscan}
collection\footnote{https://etherscan.io/}, taking the Smart Contracts bytecode, the Application Binary Interface (ABI), and the Smart Contract source code (written in Solidity) for each sample. 
The authors implemented a code parser to extract the software metrics for each smart contract considered.
They have computed: the total lines of code to a specific blockchain address, the number of smart contracts inside a single address code, blank lines,  the comment lines,  the number of static calls to events, the number of modifiers, the number of functions, number of payable functions,  the cyclomatic complexity as the simplest McCabe definition, the number of mappings to addresses for the dataset considered.
Based on the results, smart contract lines of code metric is the metric that is closest to the statistical distribution of the corresponding metric in a traditional software system.

Gencer \etal~\cite{gencer2018decentralization} proposed a measurement framework for two of the dominant cryptocurrencies, Bitcoin and Ethereum. They estimated the network resources of nodes and the interconnection among them, the protocol requirements influencing the operation of nodes, and the robustness of the two systems against attacks to analyze the depth of decentralization. They found that neither Bitcoin nor Ethereum has rigorously better properties than the other.

Other papers have computed more specific metrics tailored for specific application domains, such as healthcare~\cite{zhang2017metrics},
and model the real-time predictive delivery performance in supply chain management~\cite{meng2018blockchain}.

With respect to these metrics, the ones of the suite proposed in this paper aim at evaluating Solidity code's patterns that deteriorate the performance of the smart contract's deployment.  

\subsection{Weaknesses on Smart Contract's source code}
A part of the literature on BOS concerns the analysis and detection of different types of weaknesses affecting the code of a smart contract.


Ye \etal~\cite{ye2019towards} realized a comparison of the state-of-art bug detection tools and executed experiments to find their advantages and disadvantages.  

By analyzing the literature, Demir \etal~\cite{demir2019security} identified the vulnerabilities that developers must fix when writing smart contracts. In addition, they analyzed applications that seek these vulnerabilities and provided an overview of how they are used and which they cover. In their analysis, they identified issues related to smart contracts and provided a discussion about the problems, the challenges and the techniques of the available technology in this area.

Peng \etal~\cite{peng2019sif} proposed SIF, a framework useful for Solidity contract monitoring, instrumenting, and code generation. SIF is able to detect bugs, analyze, optimize and generate code to support developers and testers. This framework has been tested on real smart contracts deployed on the Ethereum platform.

Tikhomirov \etal~\cite{tikhomirov2018smartcheck} 
propose a classification of problems that may occur in smart contract code. Subsequently, they implemented a static analysis tool that can identify them, called Smartcheck. The tool converts Solidity source code into an XML-based intermediate representation and compares it with XPath models. The tool has been tested on a large set of real smart contracts.

Kalra \etal~\cite{dhawan2017analyzing} presented ZEUS, a framework to check the correctness and confirm the fairness of smart contracts.
By correctness, they mean using secure programming practices, instead, fairness indicates compliance with high-level business logic.
ZEUS simultaneously uses abstract interpretation, the symbolic model checking and horn clauses to speedily check the security of smart contracts. Zeus has been tested on over 22,000 smart contracts and it has shown that around 94.6\% of them are vulnerable.

In our study, we focus on the \textit{cost smells} which, at the best knowledge of the authors, is a novel area of investigation in the literature regarding BOS.

\subsection{Gas Cost Evaluation}
Literature studying the relationship between the smart contract's code and the cost of its deployment and execution has yet a small number of contributions.

\major{Baird \etal~\cite{baird2019economics} explore the economics of smart contracts.}
There is a disparity that continues to increase between the actual costs of executing smart contracts and the computational costs. 
\major{This occurs because the gas cost model of the Ethereum Virtual Machine}
(EVM) instruction-set is poorly implemented. 
To resolve this problem momentarily, the Ethereum community increases the cost of gas periodically. In their study, the authors showed a new gas cost model that fixes the principal irregularity of the current Ethereum gas cost model. Their new cost model blocks the ongoing inflation of execution time per unit of gas.

\major{Chen \etal~\cite{chen2017under} proposed analysis to show that many smart contracts}
have dependencies on the cost of gas and could be replaced by more efficient bytecodes to save gas. For this goal, they implement GASPER to automatically discover gas-costly programming patterns from the bytecode of smart contracts. Their analysis focuses on dead codes, Opaque Predicates, Expensive Operations in a Loop in relation to the gas cost using bytecode; our metrics extract information from a smart contract but at a higher level than bytecode, i.e. at the source code level.

\revise{More recently, Albert \etal~\cite{GASTAP, GASOL} proposed methods and tools for automatically inferring gas upper bounds for smart contract functions, while Grech \etal~\cite{MadMax} presented a static analysis tool for detecting gas-focused vulnerabilities in smart contracts. 
Instead, Marchesi \etal~\cite{marchesi2020design} have identified a set of 24 design patterns that influence gas consumption in the execution of Ethereum smart contract. After classifying these design patterns into 5 categories (external transaction, storage, saving space, operations and miscellaneous), for each pattern, they describe the problem and a possible solution. We can observe similarities between the cost smells and patterns identified by Marchesi \etal. In particular, based on the descriptions provided, 14 cost smells identified in our study are similar to patterns presented in prior work. \major{In contrast, our study reports five cost smells, two related to the number of loops present in the contract and the number of variables declared in them (\ie CS11 and CS1, respectively), two concerning the use of memory type arrays and public members (\ie CS7 and CS2, respectively), and one pertaining to the number of indexed parameters within events (\ie CS13), that were not discussed in~\cite{marchesi2020design}.} \major{In addition}, differently from this previous study, we (i) \change{assessed} the relevance of the identified cost smells through a survey involving real smart contract developers, and (ii) defined a suite of metrics for more easily identifying the code smells that can negatively impact gas consumption.
}
\major{On static profiling and optimization of Ethereum smart contracts, Correas \etal~\cite{correas2021static} presented a novel
static profiling technique based on static resource analysis to generate upper-bound expressions on a variety of resources (\eg the number of storage instructions, gas consumed by some EVM operations, total ether sent by an external call, etc.) that can be used for optimizing the gas consumption of smart contracts. Moreover, through the use of an automatic optimization of Solidity programs, they propose to reduce gas consumption by replacing the accesses to state variables by gas-efficient accesses to local variables. Brandstatter \etal~\cite{brandstatter2020characterizing} presented a python solidity-optimizer based on classical code efficiency optimization strategies in the context of smart contracts, postulated in
early work by Bentley~\cite{bentley1982writing} and grouped into six categories (time-for-space rules, space-for-time rules, loop rules, logic rules, procedure rules, and expression rules).
In~\cite{chen2020gaschecker}, Chen \etal proposed ten gas-inefficient programming patterns belonging to four categories. The first category regards useless code (i.e., opaque predicates and dead code). The second category encompasses the expensive operations in loops
(such as the cost smell CS1 identified in our study), the fusible loops, and the repeated computations in loops. The third category
deals with the wasted disk space (\eg storage that is never used after definition). Finally, the fourth category comprises gas-inefficient operation sequences (\ie consecutive EVM
operation sequences that can be replaced with gas-efficient
operation sequences). The authors also presented GasChecker, a tool able to detect the defined gas-inefficient patterns in the bytecode of Ethereum smart contracts. Unlike the GasMet tool that collects metrics at the source code level, GasChecker is based on symbolic execution and works at the bytecode level. }

In a different effort, Chen \etal~\cite{chen2020defining} defined 20 types of \textit{contract defects}, by analyzing posts on Ethereum StackExchange\footnote{\url{https://ethereum.stackexchange.com/}}. Among the defects identified, there are functions and data types that can increase gas consumption. 
As in our study, to validate the elicited \textit{contract defects}, they conduct a survey with practitioners and evaluate the diffusion of such defects in 587 real world smart contracts. 
However, while previous work mainly focuses on characterizing different types of defects and only suggests the implementation of practical tools for practitioners, our study has a special focus on bad coding practices that can negatively affect gas consumption, also providing developers with a concrete suite of metrics for more easily evaluating smart contract code quality from the gas consumption perspective.

\section{Elicitation of cost smells}
\label{sec:elicitation}

\begin{table*}[t!]
\centering
\caption{The list of the identified Cost Smells along with the rationale explaining the relationship with the gas consumption.}
\scriptsize
\label{tab:smells}
\resizebox{0.86\textwidth}{!}{%
\begin{tabular}{|c|c|m{2cm}|m{10cm}|c|} 
\hline
\textbf{Id}& \major{\textbf{Type}} & \textbf{Cost Smell} & \textbf{Description}  & \textbf{Ref.} \\ \hline
CS1 & \major{storage} & Duplicate writes & \revised{Modifying a variable's value several times could require much gas. To save gas, developers should overwrite variables outside cycles as much as possible.} & \cite{forum1}\\ \hline
CS4 & \major{storage} &Inefficient initialization of variables & 
An uninitialized variable is \revised{automatically} set with \revised{its} default value \revised{(\textit{e.g.,} a \texttt{uint256} variable, when not initialized, it is assumed to have the default value 0)}. \revised{When declaring a variable, explicitly setting it with its} default value \revised{is useless and wastes gas}.
& \cite{forum4}\\ \hline
CS7 & \major{storage} & Inefficient use of memory arrays  & 
Whenever a developer has to make some internal computation in a Solidity function with the help of an array, it may be good to avoid using storage, by employing \texttt{memory} arrays. If the size of the array is exactly known, fixed size \texttt{memory} arrays can be used to save gas. & \cite{forum5}\\ \hline
CS8 & \major{storage} & Inefficient use of strings & 
Using \texttt{bytes32} is cheaper than using \revised{ the }\texttt{string} \revised{type}.
\revised{If the length of the string can be limited to a certain number of bytes,} bytes1 \revised{to bytes32} data types are preferable wherever possible.
& \cite{forum6}\\ \hline
CS9 & \major{storage} & Inefficient use of return values & 
A simple optimization in Solidity consists of naming the return value of a function. It is not needed to create a local variable then. & \cite{forum7}\\ \hline
CS10 & \major{storage} & Inefficient use of global variables & 
Storing global variables in memory is expensive in terms of gas. 
\revised{Number and size} of global variables should be minimized.
& \cite{forum8}\\ \hline
CS11 & \major{storage} & Unbounded loops & 
In general, loops should be avoided. If avoiding loops is not possible, it could be beneficial to try to avoid unbounded loops, \ie loops where the upper limit of iterations is not defined.  
& \cite{forum9}\\ \hline
CS12 & \major{storage} & Inefficient use of data types & 
Use \texttt{bytes32} whenever possible, because it is the most optimized storage type. For example, storing \revised{a small number in a \texttt{uint8} variable} is not cheaper than storing it into \revised{a \texttt{uint256} variable, as, for storing, any smaller data is padded with zeros to fill the 32 bytes, requiring additional operations from the EVM and additional gas.} 
& \cite{forum9}\\ \hline
CS13 & \major{storage} & Inefficient use of indexed parameters & 
The indexed parameters in events have additional gas costs. 
\revised{It is preferable to only use the indexed qualifier for event parameters that should be searchable.}
& \cite{badr2018blockchain}\\ \hline
CS14 & \major{storage} & Inefficient use of structs & 
Since many DApps use storage, it would be useful to reduce archiving costs in order to optimize gas costs. 
\major{In particular, instance or struct variables can be packed together to reduce storage costs, while mappings can not. Thus, using a high number of mappings could result in higher storage costs than using variables that can be packed into single storage slots.}
& \cite{forum10} \\ \hline
CS15 & \major{storage} & Inefficient use of mappings & 
\major{As arrays are not stored sequentially in memory and each access to array elements requires a key-value lookup, in Solidity, arrays are more expensive versions of mappings with added features making them array-like (\eg length, bound checking, sophisticated packing behaviors, automatic zeroing out of unused storage slots, special optimizations). To save gas, mappings are preferable to arrays.}
& \cite{forum3} \\ \hline
CS17 & \major{storage} & Inefficient use of booleans & \revised{
Booleans (bool) are uint8 which means they use 8 bits of storage even if they can have only two values: True or False. When EVM packs the bools normally it can store only 32 bools in one memory slot. Otherwise, a set of 256 different booleans could be more efficiently packed in a single word by not declaring them as bool but uint256, using one bit for each boolean value.}
& \cite{forum4} \\ \hline
CS18 & \major{storage} & Inefficient use of events & 
It is cheaper to store data \revised{that is not required on-chain} in events rather than variables. 
& \cite{forum4} \\ \hline
CS2 & \major{function} & Abundance of public members & 
The order of the functions \revised{influences} the gas consumption. Since the order of the functions is based on the method ID, this implies that the subsequent ordering can consume additional gas. Depending on the VM transaction, each position will have an additional gas fee. Since all \texttt{public} members participate in the sorting, reducing public members could save gas.
& \cite{forum2}\\ \hline
CS3 & \major{function} & Scarcity of external functions  & 
Storing the input parameters in memory costs gas. \revised{For all public functions, the input parameters are copied to memory automatically. If a function is only called externally, it should be explicitly marked as \texttt{external}, in a way that these parameters are not stored into memory but are read from call data directly. } 
This can save gas when the function input parameters are huge.
& \cite{forum3}\\ \hline
CS5 & \major{function} & Inefficient use of libraries & 
\major{The bytecode of library functions is not made part of a deployed client smart contract.
Thus, smart contract developers could use software libraries to implement complex logic. Library imports help to keep the size of the client smart contract small, consequently reducing the gas required for deploying it.}
& \cite{forum4}\\ \hline
CS6 & \major{function} & Inefficient use of internal functions & 
\revised{From inside a smart contract, calls to internal functions are cheaper than calls to public functions.}
\revised{A call to a \texttt{public} function implies that all the parameters are copied into memory and passed to that function.} \revised{Conversely, a call to an internal function does not entail copying such parameters into memory again.}. \revised{For this reason, the use of \texttt{internal} functions is preferable whenever possible, } especially when the parameters are big. 
& \cite{forum4}\\ \hline
CS16 & \major{function} & Inefficient use of external calls &
Every call to an external contract costs a decent amount of gas. For optimizing gas usage, it is better to call one function and have it return all the needed data rather than calling a separate function for every piece of data.
& \cite{forum3} \\ \hline
CS19 & \major{function} & Inefficient use of functions & 
\revised{In Solidity, it could be preferable to use fewer larger functions, rather than implementing multiple functions, each performing a single small task. Indeed, multiple smaller functions cost more gas and require more bytecode.} 
& \cite{forum4} \\ \hline
\end{tabular}
}
\end{table*}

We identified a set of \textit{cost smells}, by inspecting specialized forums and books dealing with the development of Solidity smart contracts. \revised{Although exhaustively identifying any possible cost-related defect in smart contracts is not an objective of this paper, we relied on multiple sources to spot out some common bad practices in smart contract development with Solidity that can negatively impact gas consumption. In particular, we (i) leveraged some textbooks~\cite{badr2018blockchain}, highlighting good practices in smart contracts and decentralized applications development, and (ii) consulted technical blog posts dealing with Solidity and gas optimization tips. Specifically, to mine potential smell-related posts, the keywords (i) ``Solidity gas consumption", (ii) ``Solidity gas usage", (iii) ``Solidity gas optimization", and (iv) ``Solidity gas saving" have been used to perform the Google search.  By inspecting the ten top-ranked retrieved documents for each search query, we were able to discover ten blog posts specifically targeting Solidity coding practices for optimizing gas usage~\cite{forum1, forum2, forum3, forum4, forum5, forum6, forum7, forum8, forum9, forum10}.
Relying on the aforementioned information sources, we found that, in general, gas consumption could be negatively affected by:
\begin{itemize}
    \item \textit{inefficient use of data storage;}
    \item \textit{inefficient implementation of functions.}
\end{itemize}

In particular, we identified 19 \textit{cost smells} (\ie coding practices entailing higher gas consumption). 
}
Table \ref{tab:smells} reports the identified \textit{cost smells}, along with a brief description of each smell and a reference to the webpage from  which it has been deduced. In particular, the cost smells reported in Table \ref{tab:smells} could be related to either inefficient use of data storage (\ie \textit{CS1, CS4, CS7, CS8, CS9, CS10, CS11, CS12, CS13, CS14, CS15, CS17,} and \textit{CS18}) or inefficient implementation of functions (\ie \textit{CS2, CS3, CS5, CS6, CS16,} and \textit{CS19}).

\section{Study on the relevance of cost smells }
\label{sec:studyRelevance}
The \textit{goal} of this study is to investigate whether the identified cost smells (see Section \ref{sec:elicitation}) are considered as relevant by smart contract developers. To pursue this goal, we pose our first research question:

\smallskip
\emph{RQ$_1$: To what extent are the identified cost smells relevant for smart contract developers?}
\smallskip

\subsection{Research method}

To address RQ$_1$, \change{\major{similar to previous work}
on code smells assessment~\cite{
DBLP:conf/msr/BorrelliNLCP20, DBLP:conf/kbse/TufanoPBPOLP16, DBLP:journals/ese/AnicheBTGD18},} we conduct a survey targeting experts in the development of Solidity smart contracts and blockchain applications.
\major{Prior research~\cite{DBLP:conf/icsm/MantylaVL04, DBLP:conf/icsm/PalombaBPOL14} found that developers may have divergences of opinions about the incidence and the perceived relevance of code smells. Thus, to assess the extent to which the identified cost smells (reported in Table \ref{tab:smells}) are issues that might actually affect gas consumption, we decided to survey real smart contract developers and blockchain experts and ask them their opinions about the relevance of all the aforementioned smells. In particular, the survey with developers represents an important part of our study, as its results may help to avoid considering eventual cost smells for which developers report a low perceived relevance.}
Relevant guidelines on how to design and carry out survey studies in software engineering are provided in previous work~\cite{DBLP:journals/sigsoft/PfleegerK01, DBLP:journals/sigsoft/KitchenhamP02c, DBLP:journals/sigsoft/KitchenhamP02b, DBLP:journals/sigsoft/KitchenhamP02a, DBLP:journals/sigsoft/KitchenhamP02, DBLP:journals/sigsoft/KitchenhamP03}. On the one hand, according to such guidelines, we (i) \textit{design specific and measurable goals}, (ii) \textit{target subjects able to answer the questions posed}, (iii) \textit{group questions into topics}, and (iv) when possible, \textit{use standardized response formats}. On the other hand, we can not estimate whether the set of subjects involved in our survey is a representative subset of the target population, as we have no prior knowledge of such a population. Specifically, the survey is structured in three sections:

\begin{itemize}
\item {The first section aims at collecting demographic information about respondents. In particular, this section comprises questions about: (i) the highest education qualification; (ii) the domain in which respondents work (\textit{e.g.,} industry, academia, etc.); (iii) their role in the organization (\textit{e.g.,} developer,  project manager); (iv) the years of experience in software development; (v) more specifically, the years of experience in smart contract development, and (vi) the languages used to develop smart contracts.}
\item {The second section gathers the developers' perceived relevance about the identified Solidity code smells that may affect gas consumption.
Each smell is presented to respondents through a brief description of the problem, outlining its effects on gas usage and the optimizations to mitigate or avoid such effects. For each smell, we ask to provide a relevance score on a 5-level Likert scale~\cite{oppenheim2000questionnaire} (in which 1 corresponds to \textit{strongly disagree} and 5 corresponds to \textit{strongly agree}). 
Finally, for each question, the respondents could provide an optional open comment. }
\item {The final section asks questions about the general perceived usefulness of a suite of metrics for more easily identifying the identified cost smells while coding, \textit{i.e.,} (i) whether respondents perceive the availability of such a suite of metrics useful, and (ii) whether they would be willing to adopt it during smart contract development. Finally, we ask to provide free comments about possible Solidity code smells that were not considered in our study.}
\end{itemize}


To recruit participants, we posted a link to the questionnaire on Reddit channels related to Ethereum and Solidity development, namely
\texttt{solidity, dapps, cryptodevs, ethdev,} and \texttt{ethereum}. Along with the link to the online survey, we added a short message explaining its purpose, the estimated duration (20 min.), and our will to only use the collected data in aggregated, anonymized form. Besides, we sent the invitation for the questionnaire completion to our contacts who are working or studying blockchain-related topics in companies or academic institutions. Furthermore, we invited practitioners who regularly contribute to smart contract-related projects on GitHub.

We summarize the responses obtained in our survey in the form of diverging stacked bar charts, also discussing the open comments reported by the respondents.

\subsection{Results}
\label{sec:survey_results}

\begin{figure*}[t!]
    \centering
    \includegraphics[width=\textwidth]{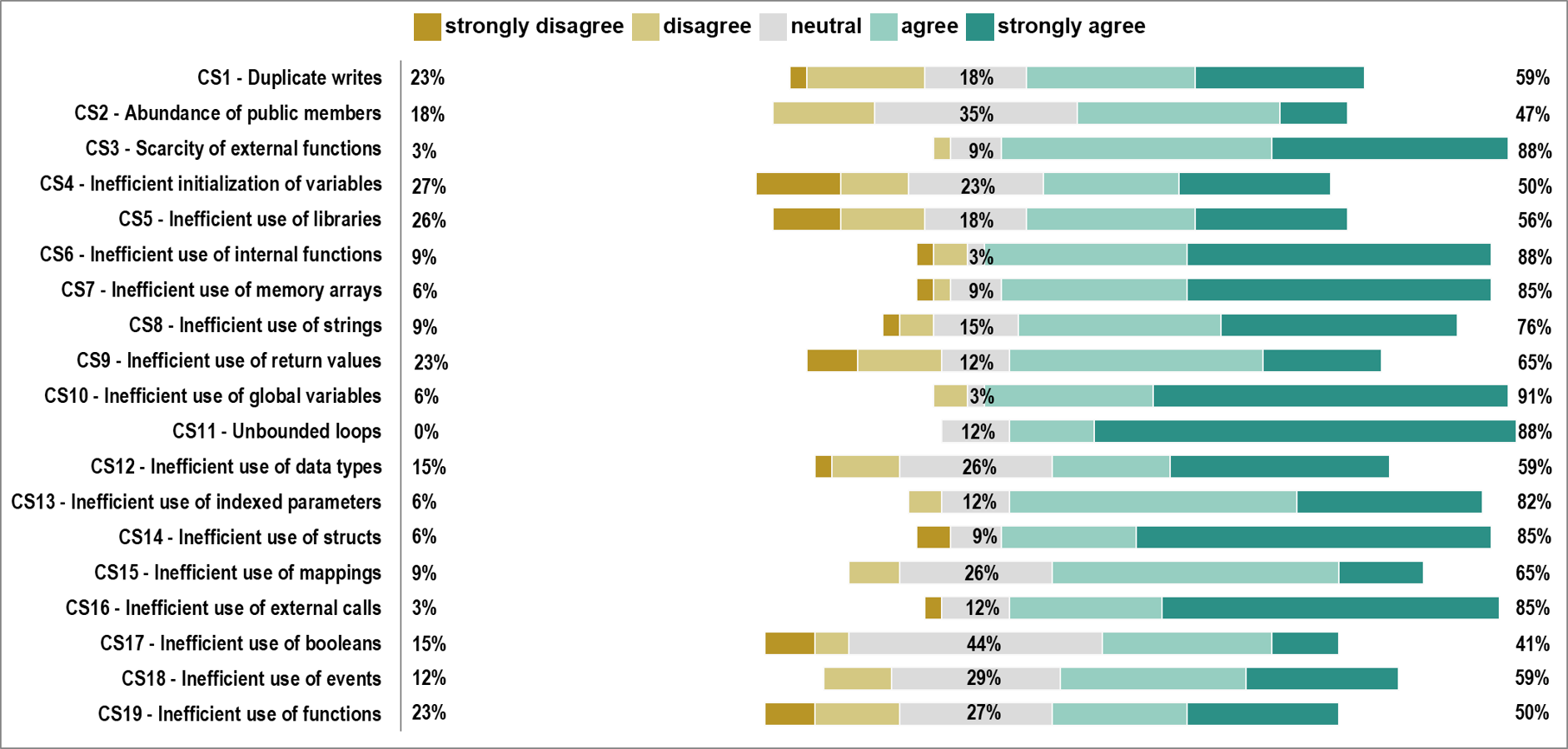}
    \caption{Perceived relevance of the 19 cost smells.}
    \label{fig:likertCostSmells}
\end{figure*}

Our survey on cost smells relevance has been kept open for one month. During this period we were able to collect responses from 34 different respondents. \change{Considering that (i) the community of smart contract developers is not so big (\eg blockchain technology is still in its infancy~\cite{grover2019diffusion}), (ii) our survey asked to assess a quite high number of different cost smells (\ie it required a moderate amount of time for completion), and (iii) the exploratory nature of our survey, we believe that 34 responses are adequate for our purposes.} 
Out of the 34 respondents, 12 own a Bachelor's degree, 12 a Master's degree, and 6 a Ph.D. 14 respondents have less than 5 years of development experience, 10 between 5 and 10 years, and 10 more than 10 years. Concerning smart contract development, 12 respondents have less than a year of experience, 16 between 1 and 3 years, and 6 more than 3 years. \change{Considering that Ethereum and Solidity were first released in late 2015 and that about 65\% of the surveyed practitioners declare more than one year of experience in smart contract development, we believe that the respondents have adequate expertise for judging the relevance of identified cost smells~\cite{chen2020defining}.}
Most of the participants (33) in our survey declare to develop smart contracts using Solidity, while just one respondent develops smart contracts with Kotlin.       
About 80\% of survey respondents (27) are smart contract developers, besides 3 blockchain researchers, 2 teachers at blockchain-related courses, 
\major{one participant involved in smart contract testing, and one (is a) product owner.}
\major{It is worth pointing out that while the majority of participants in our survey were already aware of most of the 19 smells, by looking at the survey responses, we observed that, for some smells (\ie \textit{CS2, CS9, CS12, CS13,} and \textit{CS18}),  one or two participants declared to be unfamiliar with the specific bad practices.}

In Figure \ref{fig:likertCostSmells} the diverging stacked bar charts summarize the perceived relevance of the 19 identified cost smells. The three percentages reported in the graph indicate the proportion of disagreements, neutral
responses, and agreements, respectively. 
As illustrated in Figure \ref{fig:likertCostSmells}, the majority of interviewed developers agree or strongly agree on most of the identified \textit{cost smells}. In particular, about four-fifths of the participants in our study agree or strongly agree on nine smells (\textit{i.e.,} \textit{CS3, CS6, CS7, CS8, CS10, CS11, CS13, CS14,} and \textit{CS16}), about three-fifths of them agree or strongly agree on five smells (\textit{i.e.,} \textit{CS1, CS9, CS12, CS15,} and \textit{CS18}), and for only two smells (\textit{i.e.,} \textit{CS2} and \textit{CS17}) we observe that less than 50\% of respondents agreed.  However, in these latter two cases (\textit{i.e.,} \textit{CS2} and \textit{CS17}), developers mainly tend to assume a neutral position (35\% and 44\% of respondents, respectively), rather than disagreeing. 
Indeed, concerning \textit{CS2}, surveyed developers acknowledge that public members influence gas consumption,  \emph{``though this is not the location where devs should worry too much about optimization''}. According to their opinions (i) the cost/benefit of reducing public members could be not so high (\textit{e.g.,} \emph{``The effects on gas costs due to function selector I would imagine are secondary to other savings I'd imagine''}), and (ii) the optimization should be performed at Ethereum or compiler level (\textit{e.g.,} \emph{``This should be fixed at the Ethereum or Solidity compiler level.''}). 
Similarly, for \textit{CS17}, although \emph{``bit-fiddling-concepts are an amazing source for gas-saving tactics''}, survey respondents feel that these tactics significantly hamper code readability (\textit{e.g.,} \emph{``I don't think it is worth saving gas if it impacts that much the code readability''}) and \emph{``the benefit in terms of gas consumption should be carefully evaluated''}. However, for ensuring both code readability and gas savings, they recommend that this kind of optimization should be handled at compiler level rather than source code level (\textit{e.g.,} \emph{``If this is of concern, the compiler should handle it''}). 

\begin{figure*}[t!]
    \centering
    \includegraphics[width=\textwidth]{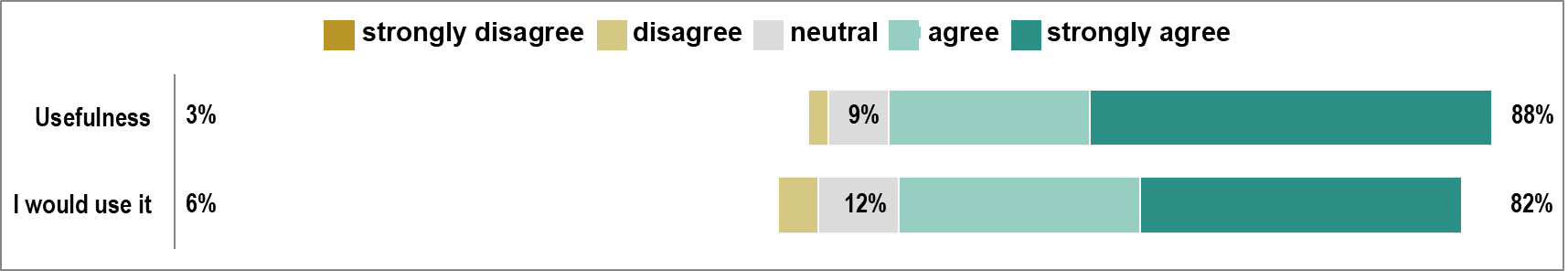}
    \caption{Survey responses on perceived usefulness of a suite of metrics for identifying cost smells}
    \label{fig:likertMetricsSuite}
\end{figure*}

For all identified smells, less than one-third of the participants disagree or strongly disagree. In the cases in which we observe higher percentages of disagreements, survey participants are mostly concerned about code readability and believe that eventual optimization should be performed at the compiler level, rather than directly in the source code. 
\major{Code readability is an important aspect to consider as developers spend a lot of time reading and inspecting code. Thus,
code readability is particularly important during smart contracts implementation and maintenance activities~\cite{iSCREAM2021}. Previous research~\cite{DBLP:conf/wcre/ChenLZ0Z21} also demonstrated that smart contract developers often reuse code blocks implemented in other smart contracts. Thus, lower readability would make the smart contract harder to understand, consequently hampering code reuse~\cite{chen2020defining}.}
As reported in previous work~\cite{zou2019smart}, it is often hard to optimize gas without reducing code readability, as more efficient code basically requires fewer instructions. This is the case of \textit{CS4}, in which, \major{even if the developers are aware that avoiding initializing a variable to its default value would be an easy strategy for gas optimization, they also argue about the negative effects on code readability (\textit{e.g.,} \emph{``It saves gas but also degrades readability''}).} Indeed, according to some of the participants in our survey, explicitly initializing variables is always a good practice, otherwise \emph{``it is difficult during a review to know if the intent for the variable was to be initialized to the default value, or if the variable is missing the initialization"}. Other participants suggest \emph{``adding a small comment to document the default value''} initialization, to save both gas and code readability. For the same reason, some of the developers involved in our study believe that default value variable initialization could be optimized at a level lower than source code (\textit{e.g.,} \emph{``Statically initializing a variable to its default value should be caught by the compiler and reduced to a no-op''}). 

Comments of a similar sort are received for \textit{CS9}, where respondents recommend balancing \emph{``the readability and the gas consumption''}, as the \emph{``benefit is minimal compared to readability''}. Indeed, they believe that the few \emph{``cost saved does not justify the decrease in code readability''}, as named return values \emph{``are frequently the source of vulnerabilities''}. Thus, since named return values help in \emph{``reducing the size of a contract,  although of a small factor''}, their usage is \emph{``an optimization the compiler should do''}. \major{This depends also on the programming style chosen (\eg \emph{``\dots it is a matter of taste whether an explicit return is better readable than naming the return variable'').}} 

Beyond code readability, developers argue that some optimizations could increase error-proneness. This is what happens for \textit{CS1}. More specifically, the survey respondents think that \emph{``it's a waste of gas to increment a storage variable in a loop''}. However, they report that the optimization reduces code quality as it is \emph{``more error-prone''}  (\textit{e.g.,} \emph{``\dots If the dev changes later the number of loop iteration, he might forget to change the increase of count, which can lead to a bug/vulnerability''}), and they also highlight that the situation where the \textit{``variable won't be used''} inside the loop  \emph{``is not common in developing's logic''}.

Instead, according to the interviewed developers, code readability is just one of the concerns related to \textit{CS5} and deriving from the adoption of libraries (\textit{e.g.,} \emph{``This has its place but can also reduce readability''}; \emph{``what is not in the contract code is not transparent on the blockchain''}). \major{Indeed, libraries can \emph{``cause all sorts of issues''} and they are rarely used \emph{``because of the difficulty of managing their deployment''}.  In addition, survey participants also highlight that the use of libraries for saving gas mainly depends on the application's context. Indeed, while using libraries could be a good strategy to reduce the deployment cost, it can also increase the execution cost (\textit{e.g.,} \emph{``With a library, you reduce the gas cost of deployment, but you increase the gas cost of executing the contract''}; \emph{``
\major{Calls to an external library (paid with each call) may become more expensive than the deployment costs (paid once)}
''}).}
Finally, the comments about \textit{CS19} indicate a general reluctance of developers to adopt coding practices that would reduce modularization (\emph{``This discourages good coding practices of modularization and small, well-contained functions''}).  In particular, the subjects in our study consider security and readability \emph{``more important than gas usage in most cases''} and keeping the business logic clear is important \emph{``for transparency and verification reasons''} (\textit{e.g.,} \emph{``Devs should aim to create small functions with a clear purpose, rather than complex functions difficult to review and to test. This is another case where the gas-saving does not justify the decrease in code readability imho''}).

Figure \ref{fig:likertMetricsSuite} reports, again in the form of diverging stacked bar charts, the results about (i) the perceived usefulness of a suite of metrics for more easily identifying the enumerated smells, and (ii) the willingness of respondents to use it if available. 30 out of 34 respondents (88\%) agreed or strongly agreed about the usefulness of a suite of metrics capturing gas consumption information, and 28 participants (82\%) indicated that they would use it in the smart contract development process. However, as highlighted by one of the survey participants the \emph{``problem of metrics is that they may become quickly outdated when the compiler changes the way it generates bytecode''}, and optimizing code for gas consumption at the bytecode level would be of broader applicability. However, since smart contract developers desire to optimize source code rather than bytecode~\cite{zou2019smart}, software metrics can allow them to directly identify the portions of source code that would have room for efficiency improvement.

\begin{center}
\fcolorbox{black}{lightgray}{
\begin{minipage}[center]{0.97\linewidth}
\textbf{RQ$_1$ Summary:} \em{Smart contract developers generally agree or strongly agree about the identified cost smells. However, they point out that some optimizations may negatively impact code readability. The vast majority of respondents are in favor of adopting a suite of source code metrics for more easily identifying cost smells.} 
\end{minipage}}
\end{center}

\section{\major{The GasMet suite and tool}}
\label{sec:suite}
\begin{table*}[t!]
\scriptsize
\centering
\caption{\major{Metrics' descriptions and correlations.}}
\begin{tabular}
{|m{3.5cm}|m{0.5cm}|m{11cm}|m{0.5cm}|m{0.5cm}|}
\hline 
\textbf{Metric's name} &  \textbf{Abbr.} & \textbf{Description}& \textbf{CS}  &\textbf{Corr.}\\ \hline
Assignments  within  Cycles & ACI &  It computes the assignments and/or variable updates occurring within loops. & CS1 & P \\ \hline 
Public Members & PM & It enumerates the functions defined as public members. & CS2 & P\\ \hline 
External Functions & EF & It enumerates the functions defined as external. &  CS3 & N \\ \hline 
Assignments to default values & AZ & It computes the assignments to default values during variable definitions. & CS4 & P \\ \hline 
Library Imports & LI & It estimates the usage of external libraries.
Given the number of \textit{import} statements, $S_{import}$: $$LI = S_{import}$$
& CS5 & N \\ \hline 
Internal Functions & IFF & It computes the ratio of \textit{internal} functions on the total number of defined functions. 
Given the number of \textit{internal} functions, $F_{internal}$, and the number of overall functions, $F_{all}$, defined in the contract: $$IFF  = \frac{F_{internal}}{F_{all}}$$
& CS6 & N \\ \hline
Uses of Memory Arrays & UMA & It computes the ratio of \textit{memory} arrays on the total \major{number} of defined arrays within the contracts. Given the number of arrays defined as \textit{memory}, $A_{memory}$, and the total \major{number} of defined arrays, $A_{all}$: $$UMA = \frac{A_{memory}}{A_{all}}$$ & CS7 & N \\ \hline

Strings and Bytes & SB & It computes the occurrences of \textit{string} variables with respect to the occurrences of \textit{bytes}. Given the number of \textit{string} variables, $V_{string}$, and the number of \textit{bytes} variables, $V_{bytes}$: $$SB = \frac{V_{string}}{(V_{bytes} +  \major{V_{string}})}$$ & CS8 & P \\ \hline
Functions Returning Local Variables & RLV & It computes the ratio of functions returning local variables on the total number of functions. Given the number of functions returning local variables, $F_{local}$, and the overall number of functions, $F_{all}$, defined in the contract: $$RLV = \frac{F_{local}}{F_{all}}$$ & CS9 & P
\\ \hline

Global Variables & GV & It computes the number of global variables.
& CS10 & P \\ \hline
Number of Loops & NLF 
& It computes the number of loops inside the contract. 
\major{In the case of a code block with two (or more) nested loops, the NLF metric will count two (or more) loops.}
& CS11 & P \\ \hline
Number of non-32-bytes variables & NU & It computes the ratio of non-32-bytes type variables on the total number of variables. & CS12 & P \\ \hline
Indexed Parameters & IP & It computes the number of parameters declared as \textit{indexed} within events. & CS13 & P \\ \hline 
Number of Mappings & NM & It computes the occurrences of \textit{mapping} with respect to occurrences of instance variables. & CS14 & P \\ \hline
Mappings and Arrays & MA & It computes the ratio of \textit{mappings} with respect to the sum of mappings and \textit{arrays} defined in the contract. Given the occurrences of mappings, $N_{mappings}$ and the occurrences of arrays, $N_{arrays}$: $$MA = \frac{N_{mappings}}{(N_{mappings}+N_{arrays})}$$
& CS15 & N \\ \hline

External Calls & EC & It computes the ratio of calls to external functions (\ie not defined in the contract) on the total number of function calls. Given the occurrences of calls to external functions, $C_{external}$, and the total number of calls within the contract,  $C_{all}$: $$EC = \frac{C_{external}}{C_{all}}$$ & CS16 & P
\\ \hline
Boolean Variables & BV & It computes the ratio of \textit{boolean} variables on the number of total variables. Given the number of variables of the boolean type, $V_{bool}$, the number of overall variables, $V_{all}$: $$BV = \frac{V_{bool}}{V_{all}}$$
& CS17 & N
\\ \hline
Number of Events & NE & It computes the number of \textit{events}. 
& CS18 & N
\\ \hline
Defined Functions & DF & It computes the number of the functions with respect to the overall lines of code. Given the number of overall defined functions, $F_{all}$, and the amount of lines of code, $LOC$: $$DF = \frac{F_{all}}{LOC}$$ & CS19 & P \\ \hline

\end{tabular}
\label{tab:Metric correlation}
\end{table*}


Relying on the cost smells reported in Table \ref{tab:smells}, we define a suite of metrics \major{(detailed in Table \ref{tab:Metric correlation})}, called GasMet metrics, able to capture information related to each smell. \major{Table~\ref{tab:Metric correlation} reports, for each defined metric, its name, its acronym, its description, the cost smell to which it is related, and the expected direction of the correlation between the metric and the gas consumption. An expected positive (P) correlation between the metric and the gas consumption means that higher values of the metric should correspond to higher gas consumption. On the contrary, an expected negative (N) correlation between the metric and the gas consumption means that higher values of the metric should correspond to lower gas consumption.}

\major{It is worth pointing out that some metrics (\eg DF) consider the lines of code of the smart contract, while some others (\eg PM) do not. This choice relies on the expected impact on gas consumption. For example, according to CS2, ten public members would have the same impact on the gas consumption independently of the smart contract's lines of code. Contrarily, DF is a proxy to measure code modularization, and, according to CS19, ten defined functions would have different gas consumption impacts in smart contracts with different sizes.}

The defined metrics could help in statically measuring the code quality of smart contracts, from the gas consumption perspective. 

They could represent a guide for developers for improving this facet of smart contract's quality, as very localized changes might be required to improve the values of the specific metrics. For instance, to achieve lower values of the \textit{ACI} metric (and, consequently, save gas), it is sufficient to reduce the number of variable updates occurring within cycles.
\major{More specifically, the defined metrics can help developers detecting (i) gas-inefficient or redundant operations (\eg ACI, AZ, NLF, RLV), (ii) function visibility (and responsibility) inefficiencies (\eg PM, EF, IFF), (iii) inefficient usages of data types and structures (\eg UMA, SB, GV, NU, IP, NM, MA, BV, NE), and (iv) inefficiencies in code modularity  (\eg LI, EC, DF). For example, the function visibility-related metrics can help developers identifying likely opportunities for optimizations (\eg they could change the modifier of functions that are only used internally from \texttt{public} to \texttt{internal}, or update the modifier of functions that are not used internally from \texttt{public} to \texttt{external}). Similarly, as events consume less gas than Solidity variables, the number of events (in combination with the number of variables) represents a good indicator for developers interested in applying optimizations. Indeed, by applying updates to the smart contract's logic, they could identify data (not required on-chain) to store in events and save gas.}

\begin{figure*}[t!]
    \centering
    \includegraphics[width=0.90\textwidth]{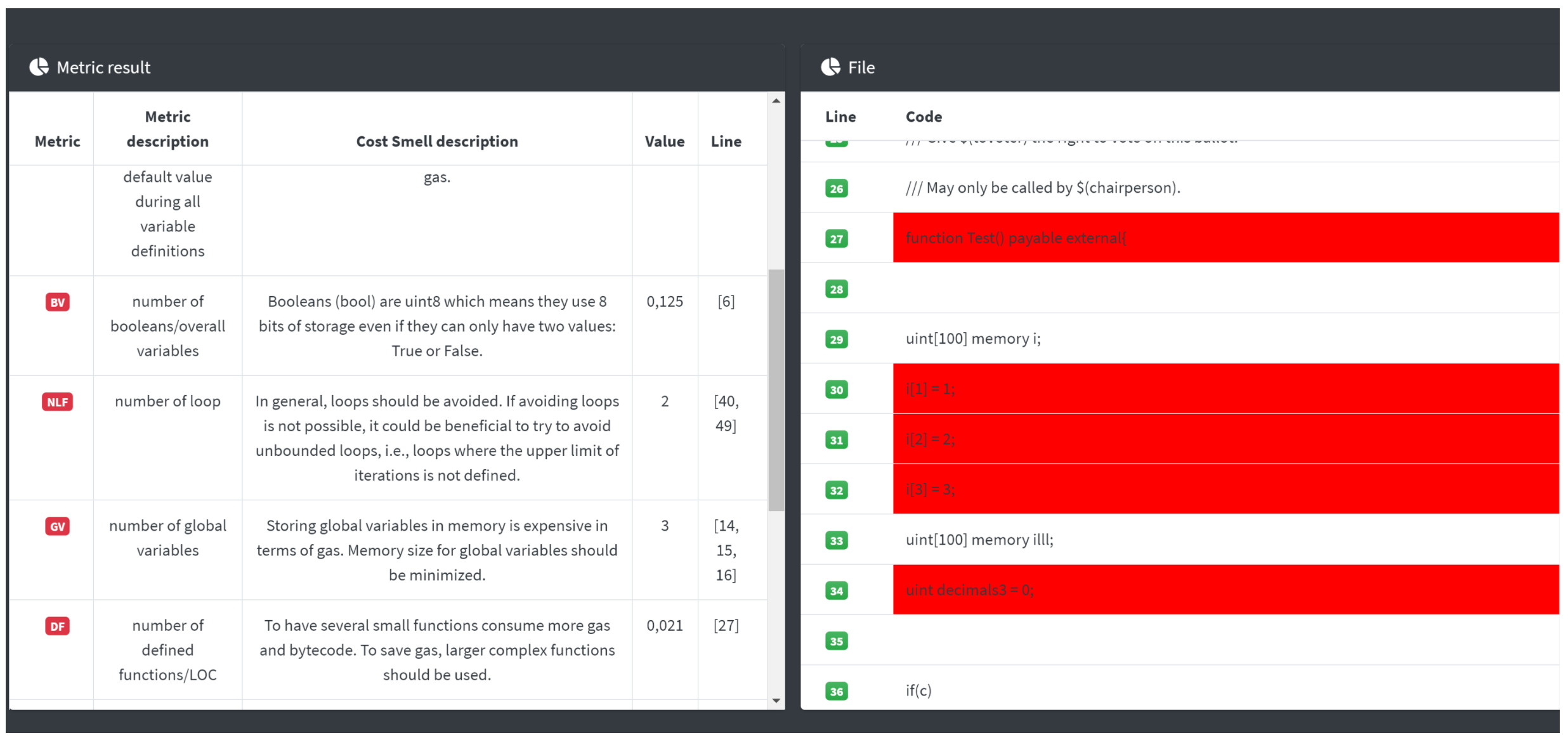}
    \caption{GasMet tool's usage example.}
    \label{fig:metricResult}
\end{figure*}

\major{For helping smart contract developers more easily identifying likely gas-inefficient code portions in smart contracts written in Solidity, we implemented the GasMet tool.}
\major{GasMet is a prototype Java tool able to parse Solidity smart contracts and automatically compute the metrics in the GasMet suite. In particular, for properly parsing smart contracts, 
the tool leverages a generated parser based on a modified version of an existing \texttt{antlr4}  grammar\footnote{\url{https://github.com/solidityj/solidity-antlr4}}. 
Specifically, the tool is composed of three main software modules: the \texttt{Lexer}, the \texttt{Parser}, and the \texttt{GasMet Metrics Calculator}. 
The \texttt{Lexer} receives as input the raw text of the Solidity smart contract and provides a stream of tokens, relying on lexical rules. The \texttt{Parser} processes this stream and builds an abstract syntax tree. The \texttt{GasMet Metrics Calculator} traverses the tree generated by the \texttt{Parser} to compute the GasMet metrics.
Once processed the smart contract, the \texttt{GasMet Metrics Calculator} stores the results in a tabular format. The results can also be exported as a CSV file. 



}

\major{
The tool either provides a graphical user interface (GUI) or can be used from the command line. The instructions for running the GUI or using it from the command line are provided in our replication package\footnote{\url{https://github.com/paperSubmition2020/GasmetReplicationPackage/tree/master/GasMetSuite}}. 
In particular, the tool's GUI is built as a web application that can be deployed by using the Wildfly application server.
Once selected a Solidity smart contract to analyze, the tool's GUI presents two frames (see Figure \ref{fig:metricResult}). The left frame reports a table containing the results of the GasMet metrics computation on the selected smart contract. More specifically, the aforementioned table encompasses the following information about each metric: acronym, extended name, description, computed value, and lines involved in the computation. In the right frame, the source code of the smart contract under analysis is shown. The lines involved in the computation of the different GasMet metrics are highlighted in red to allow developers and researchers more easily identifying the code portions that could likely require improvements. A developer interested in saving the gas required for the deployment of her smart contract could analyze the smart contract through the GasMet tool and apply improvements to achieve better values for the computed metrics.
There are many different tools available for Ethereum smart contracts. These tools have been gathered from research publications and through Internet searches. To make a comparison with the GasMet tool, we selected the tools that are actively maintained, open-sourced, ready for use, such as Remix-IDE\footnote{https://remix-ide.readthedocs.io/en/latest/} and SmartCheck\footnote{https://github.com/smartdec/smartcheck}.
Unlike our tool, Remix-IDE is a browser-based IDE for developing Solidity contracts. It can connect to the Ethereum network using Metamask\footnote{https://metamask.io/} and developers can directly deploy smart contracts from Remix. During compilation it is able to report security issues, indicating where they occurred in the code. It also reports implicit visibility, unchecked return values, implicit typing, deprecated constructs, and address checksums. The static analysis is only lightweight and includes some control flow analysis. Remix-IDE also enables the testing of smart contracts via unit tests written using tape\footnote{https://www.npmjs.com/package/tape}. Remix-IDE, for the Solidity static analysis, is not based on a suite of metrics but computes the gas consumption associated with each Ethereum Assembly instruction listed in the Ethereum Yellow-Paper/AppendixG\footnote{https://ethereum.github.io/yellowpaper/paper.pdf}. 
SmartCheck is a static analysis tool for smart contracts written in Solidity and Vyper. It is developed by SmartDec and the University of Luxembourg. Like other static analysis tools, it works at the source code level. In particular, it transforms the source code into an  XML-based intermediate representation. This representation is then checked against XPath patterns to highlight potential vulnerabilities in the code.


}


\section{Study on the relations between GasMet metrics and deployment costs}
\label{sec:design}
The \textit{goal} of this second study is to more-in-depth investigate the relationships between the individual metrics in the \textsc{GasMet} suite and smart contracts' deployment costs. More specifically, this investigation aims at assessing whether our suite can capture gas-related information by analyzing smart contracts' source code. To pursue this goal, we pose our second research question:

\smallskip            
\emph{RQ$_2$: To which extent does the GasMet suite correlate with gas consumption?}
\smallskip

\subsection{Context selection and data extraction}
\label{sec:dataCollection}

For this study, we collected a dataset \change{containing the source code of} 2,186 Solidity smart contracts \change{deployed on Ethereum.} 
\change{Actually, there are about 1.5 million smart contracts deployed on Ethereum~\cite{DBLP:journals/fi/PierroTM20}. However, for many of these smart contracts, the source code is not publicly available~\cite{DBLP:conf/uss/ZhouKBMMB18}. For this reason, we leveraged Etherscan\footnote{\url{https://etherscan.io/}}, a popular service for Ethereum blockchain exploration that offers a feature called \emph{``verified contracts”}, through which developers can publish the source code of blockchain smart contracts. The Etherscan API actually allows obtaining the source code of more than 40,000 smart contracts~\cite{DBLP:journals/ese/OlivaHJ20}. As we needed to compile and deploy the extracted contracts to estimate the deployment costs and this requires a discrete amount of time, for reducing the experimentation time, we decided to conduct our study on a (statistically significant) set of contracts randomly sampled from the initial collection returned by Etherscan. It is worth noticing that the number of instances in our dataset was defined for guaranteeing high representativeness. \major{Indeed, our dataset, beyond being a statistically significant sample of the Etherscan collection (\ie a confidence level of 99\% and a margin of error smaller than 3\%), it is also a statistically representative sample of all the smart contracts deployed on Ethereum (\ie a confidence level of 99\% and a margin of error smaller than 3\%).} 
It is worth noticing that the margin of error was computed 
according to the formula for margin of error with finite population correction~\cite{salkind2010encyclopedia}:

$$Margin~of~error = z * \sqrt{\frac{p * (1 - p)}{(N-1)*\frac{n}{N-n}}}$$

where $z$ is the z-score associated with the confidence level ($z = 2.576$ in the case of a confidence level of 99\%), $p$ is the sample proportion ($p = 0.5$ in the case of random sampling), $n$ is the sample size ($n = 2186$ in our case), and $N$ is the population size.
}

For each smart contract in our dataset, we extracted 
the values of the GasMet metrics, by using the parser we developed \major{(see Section \ref{sec:suite})}.

Table \ref{tab:SLOC} groups the smart contracts considered in our data set according to the number of lines of source code (SLOC). It is worth noticing that about 73\% of the analyzed contracts have a size expressed in lines of source code between 50 and 500, while only a limited number of them (i.e., about 6\%) exhibit higher values of SLOC.

\begin{table}[]
\centering
\caption{Lines of source code of the analyzed contracts.}
\begin{tabular}{|c|c|c|}
\cline{1-3}
\textbf{\# of source code lines} & \textbf{\#instance} & \textbf{\%instance}  \\ \cline{1-3}
SLOC \textless 50                              & 462                             & 21.1 \\\cline{1-3}
50 $\leq$ SLOC $<$ 100                         & 800                             & 36.6 \\\cline{1-3}
100 $\leq$ SLOC $<$ 500                        & 794                             & 36.3  \\\cline{1-3}
SLOC $\geq$ 500                                    & 130                      & 5.9 \\\cline{1-3}
\end{tabular}
\label{tab:SLOC}
\end{table}
To collect data on gas consumption, we used the built-in smart contract compilation. The compilation process for smart contracts involves the Truffle suite~\cite{truffle} and the Ethereum client Ganache~\cite{ganache} where it gets deployed.
Truffle Suite is a collection of tools for the development and testing of Ethereum blockchain based software. It contains Truffle which is the most popular development and testing framework using the Ethereum Virtual Machine (EVM). 
The compilation and deployment pipeline of Ethereum smart contract, that we used, is as follows:
\begin{enumerate}
\item \textit{setting up an Ethereum development environment}: we create a Truffle project, read smart contracts from dataset, create a deployment script that deploys and initializes the state of deployed contracts on blockchain specified in the project config file, and
\item \textit{collect data regarding gas consumption}: Ganache is a local test blockchain included in the above mentioned Truffle Suite. The gas consumption, on local blockchain, is computed by: \textit{gasCost} * \textit{gasPrice}. \textit{gasPrice} represents the price to pay per gas unit to deploy the contracts under Truffle project and was set, in our experimentation, to the value of 1 Wei (1 Ether = \(10 ^{18}\) Wei).
\textit{gasCost} is the maximum number of gas unit the EVM can use to process the contract deployment transaction. 

\end{enumerate}


\textbf{Replication package.} All the analyzed contracts source code (\texttt{.sol} files), as well as the metrics' results, are made publicly available in our replication package\footnote{\url{https://github.com/paperSubmition2020/GasmetReplicationPackage}}.

\subsection{Analysis method}
\label{sec:analysisMethod}

For answering RQ$_2$, we considered the values of GasMet metrics computed on the smart contracts in our dataset and the related values of gas consumption collected by deploying such smart contracts (see in Section \ref{sec:dataCollection}). \change{Similar to previous work~\cite{DBLP:journals/smr/AjienkaVC20}, we investigated the correlation between source code metrics computed on several smart contracts and the resources needed for their deployment (i.e., \textit{gasUsed}), as well as the correlations between each pair of metrics.} 
More specifically, to evaluate if significant relationships can be found between (i) each of the GasMet metrics and the gas consumption, and (ii) each pair of GasMet indicators, we used the Spearman rank correlation coefficient~\cite{daniel1990spearman}, fixing the p-value $\leq 0.05$ and adopting the Holm’s p-value correction procedure~\cite{holm1979simple} to deal with multiple comparisons. 
\major{We ran statistical significance tests, with $\alpha = 0.05$, for being sure that there is at maximum a 5\% probability that the strength of the relationship found (the $\rho$ coefficient) happened by chance (when p-value $\leq 0.05$). In particular, we tested the following null hypothesis:}  

\noindent
\major{\textit{$H_0$: There is no monotonic association between the metrics $m_i$ and $m_j$}} 

\major{where $m_i$ and $m_j$ $\in$ \{\textit{GasCost, ACI, PM, EF, AZ, IFF, UMA, SB, RLV, GV, NLF, NU, IP, NM, MA, EC, BV, NE, DF}\} and $i \neq j$.}
We interpret the strength of the correlation as (i) \textit{small} for $0.10 \le \lvert \rho \rvert \le 0.29$, (ii)  \textit{medium} for $0.30 \le \lvert \rho \rvert \le 0.49$, and (iii) \textit{large} for $\lvert \rho \rvert \geq 0.50$, as recommended by Cohen's standard~\cite{cohen1988statistical}.

To better understand which of the metrics in our suite might affect gas consumption more, we performed a Random Forest (RF) regression analysis. To perform this analysis, we considered a dataset containing the values of all the GasMet metrics computed on the 2,186 smart contracts selected for our study (see Section~\ref{sec:dataCollection}) and the corresponding values of gas consumption collected by deploying such smart contracts.
In particular, we tried to predict gas consumption (\ie dependent variable) based on the values of the GasMet metrics (\ie independent variables).
\major{In this analysis, we leveraged the \texttt{randomForest}~\cite{randomForestPackage} package from RStudio to estimate the performance of the models and the importance of variables. We used the Random Forest regression method as it works well with almost all types of data, generally does not overfit,  and it is easy to get the relative importance of the predictor variables from a trained model~\cite{DBLP:journals/ese/DeyM20}. 
Specifically, we first performed a grid search to find the best model parameters (i.e., \texttt{ntree}, the number of trees to grow, and \texttt{mtry}, the number of variables randomly sampled as candidates at each split), by using 10-fold cross-validation. The best model (i.e., the one with lowest Mean Square Error) was obtained by setting \texttt{ntree = 500} and \texttt{mtry = 8}. Then, the following steps were performed:
\begin{enumerate}
    \item We split the initial dataset into a training set $T_{train}$ (\ie 75\% of the initial dataset) and the corresponding test set $T_{test}$ (\ie 25\% of the initial dataset) at random.
    \item The RF algorithm (with \texttt{ntree = 500} and \texttt{mtry = 8}) was trained (on $T_{train}$) and tested by predicting the samples in $T_{test}$. Popular metric were used to assess the performance of the model: Mean Absolute Error (MAE), Root Mean Square Error (RMSE), and the Coefficient of determination (R$^2$)~\cite{sammut2011encyclopedia}. In particular, MAE measures the average of the residuals (\ie the absolute difference between actual and predicted values), RMSE represents the standard deviation of residuals, and R$^2$ is the proportion of the variance in the dependent variable (\ie the gas consumption) which is explained by the model.
    \item We estimated the importance of the different metrics by considering the samples in $T_{train}$ and the percentage increase in Mean Square Error (\ie \%IncMSE) of each independent predictor (\ie each GasMet metric). The \%IncMSE metric represents the deterioration of the predictive ability of the model when a predictor is randomly permuted  and the other predictors remain unchanged~\cite{DBLP:journals/bmcbi/StroblBKAZ08}. 
    In particular, each tree in a random forest has its out-of-bag (OOB) sample of data that was not used during construction. This sample is used to compute the importance of a specific variable. For each tree, the prediction error (MSE) on the out-of-bag portion of the data is recorded. Then the same is done after randomly shuffling the values related to a specific GasMet metric, keeping all other variables the same. 
    The differences between the two MSE values (\ie MSE on correct data and MSE on permuted data) obtained for each tree are then averaged over all trees and normalized according to the standard deviation of the differences~\cite{randomForestPackage}. 
    Finally, the percentage increase in MSE (\%IncMSE) on the shuffled data is measured. 
    This procedure has been applied for all the 19 GasMet metrics. The higher the \%IncMSE is, the higher the importance of the respective predictor in relation to the target variable is~\cite{sabatti2009applied}.
\end{enumerate}

To reduce sampling bias and obtain more reliable results, we repeated steps 1,2, and 3 ten times. Therefore, all the results were averaged over the ten runs.}


\subsection{ Results}
 
 \begin{figure*}[t!]
    \centering
    \includegraphics[width=0.76\textwidth]{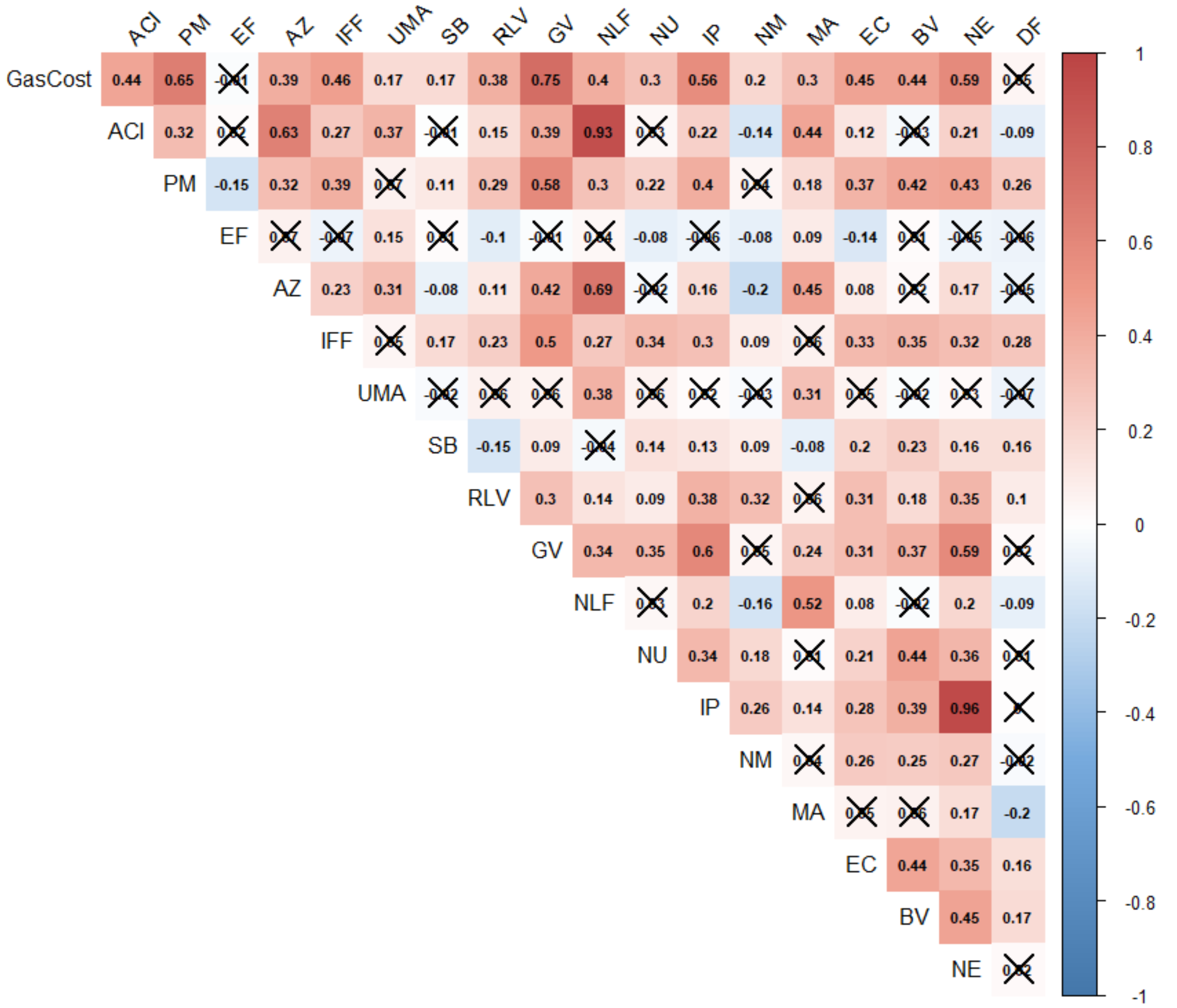}
    \caption{Spearman correlation results among the metrics belonging to the GasMet suite and the gas cost (values with x above 
    is used to indicate the correlations that are not statistically significant).}
    \label{fig:KendallGasMet}
\end{figure*}
Figure \ref{fig:KendallGasMet} reports the results of the Spearman 
\major{correlation between each pair of the metrics in the GasMet suite} and between the GasMet metrics and gas consumption. 
\major{As specified in Sections \ref{sec:dataCollection} and \ref{sec:analysisMethod}, for each smart contract in our dataset, we used the GasMet tool to compute the GasMet metrics, while we leveraged Truffle (for compiling and deploying the smart contract) and Ganache (a locally deployed blockchain simulator) to collect data about gas consumption. Once obtained the metrics and gas consumption values for all the smart contracts in our dataset, we implemented a script in the R language (using the \texttt{corrplot}\footnote{https://www.rdocumentation.org/packages/corrplot/versions/0.90} library) to compute the correlation values.}
It is worth noticing that in Figure \ref{fig:KendallGasMet} we do not consider the \textit{LI} metric, as when computing this metric on the instances in our dataset, all zero values have been obtained.

All the results considered in the following discussion correspond to an adjusted p-value $\leq$ 0.05.



Four metrics of the GasMet suite are more strongly linked to the gas consumption (in accordance with the Cohen's standard), since their correlations with the gas cost have a \textit{large} effect size, which are: (i) the occurrences of public members (\textit{PM}, with $\rho = 0.65$), (ii) the number of global variables (\textit{GV}, with $\rho = 0.75$), (iii) the number of indexed parameters (\textit{IP}, with $\rho = 0.56$), and (iv) the number of events (\textit{NE}, with $\rho = 0.59$). Higher PM values correspond to a higher gas consumption since public members are hash-sorted, and the sorting algorithm entails a certain fixed quantity of gas (depending on the EVM transactions) for each position in the list. Global variables cause an additional cost due to the storing of information within the smart contract's state on the blockchain, so their usage must be reduced and replaced with local variables which are not stored on the blockchain. Similarly, indexed parameters consume more gas than non-indexed parameters, while, if properly used, events could be adopted to store data that is not required on-chain.   



With regards to the metrics that exhibited a \textit{medium} effect size in the correlation with the gas cost, nine should be mentioned: 
(i) the ratio of internal functions (\textit{IFF}) with $\rho = 0.46$, (ii) the ratio of external calls (\textit{EC}) with $\rho = 0.45$, (iii) the ratio of boolean variables (\textit{BV}) with $\rho = 0.44$, (iv) the assignments within cycles (\textit{ACI}) with $\rho = 0.44$, (v) the number of loops (\textit{NLF}) with $\rho = 0.40$, (vi) the assignments to default values (\textit{AZ}) with $\rho = 0.39$, (vii) the number of functions returning local variables (\textit{RLV}) with $\rho = 0.38$, (viii) the number of non-32-bytes variables (\textit{NU}) with $\rho = 0.30$, and (ix) the ratio of mappings (\textit{MA}) with $\rho = 0.30$. 


\begin{table}[t!]
\scriptsize
\centering
\caption{Survey responses and correlations between GasMet metrics and gas consumption.}
\begin{tabular}{|m{0.7cm}|m{2.6cm}|m{0.7cm}|m{2.6cm}|}
\hline 

\textbf{Metric} &  \textbf{Correlation with gas consumption (Spearman's $\rho$}) & \textbf{Cost smell (CS)}& \textbf{\% of survey respondents who agreed on the CS relevance}\\ \hline
ACI & $\rho = 0.44$ & CS1 & 59\% \\ \hline 
PM & $\rho = 0.65$ & CS2 & 47\% \\ \hline
EF & not significant & CS3 & 88\% \\ \hline
AZ & $\rho = 0.39$ & CS4 & 50\% \\ \hline
LI & not significant & CS5 & 56\% \\ \hline
IFF & $\rho = 0.46$ & CS6 & 88\% \\ \hline
UMA & $\rho = 0.17$ & CS7 & 85\% \\ \hline
SB & $\rho = 0.17$ & CS8 & 76\% \\ \hline
RLV & $\rho = 0.38$ & CS9 & 65\% \\ \hline
GV & $\rho = 0.75$ & CS10 & 91\% \\ \hline
NLF & $\rho = 0.40$ & CS11 & 88\% \\ \hline
NU & $\rho = 0.30$ & CS12 & 59\% \\ \hline
IP & $\rho = 0.56$ & CS13 & 82\% \\ \hline
NM & $\rho = 0.20$ & CS14 & 85\% \\ \hline
MA & $\rho = 0.30$ & CS15 & 65\% \\ \hline
EC & $\rho = 0.45$ & CS16 & 85\% \\ \hline
BV & $\rho = 0.44$ & CS17 & 41\% \\ \hline
NE & $\rho = 0.59$ & CS18 & 59\% \\ \hline
DF & not significant & CS19 & 50\% \\ \hline
\end{tabular}
\label{tab:MetricsCScorrespondence}
\end{table}

As reported in Table~\ref{tab:MetricsCScorrespondence},
\major{among the metrics exhibiting \textit{large} relationships with gas consumption, for \textit{GV, IP,} and \textit{NE} we observe that the majority of surveyed developers (\ie $ > 58\%$) agree or strongly agree on the relevance of the related smells (\ie  \textit{CS10, CS13,} and \textit{CS18}). Only 47\% of them agree or strongly agree on the relevance of the \textit{CS2} smell (related to the \textit{PM} metric). However, as reported in Section \ref{sec:survey_results}, the developers acknowledge that the \emph{efficient usage of public members can save gas}, but they  believe that the \emph{optimizations should be performed at the Ethereum or compiler level}. Similarly, among the metrics exhibiting \textit{medium} relationships with gas consumption, we observe that, for most of them (\ie \textit{ACI, IFF, RLV, NLF, NU, MA,} and \textit{EC}), the majority of surveyed developers (\ie $ > 58\%$) agree or strongly agree on the relevance of the related smells (\ie \textit{CS1, CS6, CS9, CS11, CS12, CS15,} and \textit{CS16}). In contrast, only 50\% of them agree or strongly agree on the relevance of the \textit{CS4} smell (related to the \textit{AZ} metric), and 41\% of the participants in our survey agree or strongly agree on the relevance of the \textit{CS17} smell (related to the \textit{BV} metric). As reported in Section \ref{sec:survey_results}, although developers recognize that \textit{CS4} and \textit{CS17} smells could be sources of gas wasting, they rather worry about the risk of hampering code readability that could derive from the optimizations. Curiously, the majority of surveyed developers agree or strongly agree on the relevance of the cost smells related to metrics exhibiting \textit{small} or not significant relationships with gas consumption (\ie \textit{EF, UMA, SB, NM,} and \textit{DF}). Thus, further investigation is needed to understand whether it is possible to define better metrics for capturing the occurrences of the corresponding smells (\ie \textit{CS3, CS7, CS8, CS14,} and \textit{CS19}).}

\major{Five correlations with gas consumption have been observed to be weak or not significant (\ie UMA, SB, NM, MA, and DF). 
Using a \textit{memory} array (UMA), as it happens with any other kind of variable, means that the values of the data structure are not saved in \textit{storage}. \textit{Storage} in Solidity refers to a mechanism that holds data between function calls, while \textit{memory} keyword makes Solidity to create a chunk of space for the variable at method runtime,  guaranteeing its size and structure for future use in that method. By using a metaphor, \textit{storage} could be seen as a hard drive, while \textit{memory} as RAM. However, in our dataset we observed a very low usage of \textit{memory} arrays (for 2,112 smart contracts in our dataset the value of this metric is 0). This could explain why UMA is weakly correlated with the gas cost. Concerning the usage of \texttt{string} type instead of \texttt{bytes} type (SB), we expected different results. Consequently, we did not find quantitative evidence about this smell. Also in this case, for the majority of the smart contracts in our dataset, we observe that the value of this metric is 0. Since the results of the survey supports CS8, we can conclude that this smell deserves further investigation. 
Mappings are lightweight data structures, free of additional functions like the length computation, bound checking, and optimization. Similar to the other metrics exhibiting weak correlation with gas consumption, we observe that for many smart contracts in our dataset we report a MA value equal to 0. This could explain why MA resulted as weakly correlated with the gas cost. On the contrary, the NM metric assumes non-zero values for the great majority of smart contracts in our dataset and we believe that further investigation is needed to understand the reason behind the weak correlation between the number of mappings and the gas consumption.  The number of defined functions (DF) is not significantly correlated with the gas cost, while the proliferation of public members (PM) might affect gas consumption significantly.}

By observing the pairwise correlations between the metrics, we can discover the developers' habits in writing smart contracts that can negatively influence gas consumption.
Among the correlations with a \textit{large} effect size, we can observe (i) the one between the Number of Indexed Parameters (\textit{IP}) and Number of Events (\textit{NE}), with $\rho = 0.96$, and (ii) the one between the Assignment Within Cycles (\textit{ACI}) and Number of Loops (\textit{NLF}), with $\rho =0.93$. From these two relationships it emerges that two bad practices are pretty common: the first one is the use of indexed parameters within the events, while the second one consists of valuing variables inside the cycles. As previously mentioned, both the practices lead to an increasing of gas cost, and should be discouraged. 
The correlation between \textit{ACI} and \textit{AZ}, with $\rho = 0.63$, suggests that assignments to default values often occur within the cycles and this trend is also confirmed by the \textit{large} correlation between \textit{NLF} and \textit{AZ} ($\rho = 0.69$). The \textit{large} effect size ($\rho = 0.52$) of the correlation between the Number of Loops (\textit{NLF}) and the Mappings and Array (\textit{MA}) metric also implies that loops are frequently used to iterate over mappings. 
\major{Figure \ref{fig:KendallGasMet} also shows the correlations between the various types of functions. In particular, focusing on the metrics modeling internal (IFF), external (EF), public (PM), and defined (DF) functions, we only observe a correlation with \textit{medium} effect size ($\rho = 0.39$) between the PM-IFF pair, while for all the other pairs only small or not significant correlations occur.} 
Finally, the number of global variables (\textit{GV}) usually grows when higher numbers of (i) public members (\textit{i.e.}, \textit{PM}, $\rho = 0.58$), (ii) internal functions (\textit{i.e.}, \textit{IFF}, $\rho = 0.50$), (iii) events (\textit{i.e.}, \textit{NE}, $\rho = 0.59$), or (iv) indexed parameters (\textit{i.e.}, \textit{IP}, $\rho =0.60$) are adopted.

\begin{figure}[t!]
    \centering
    \includegraphics[width=0.48\textwidth]{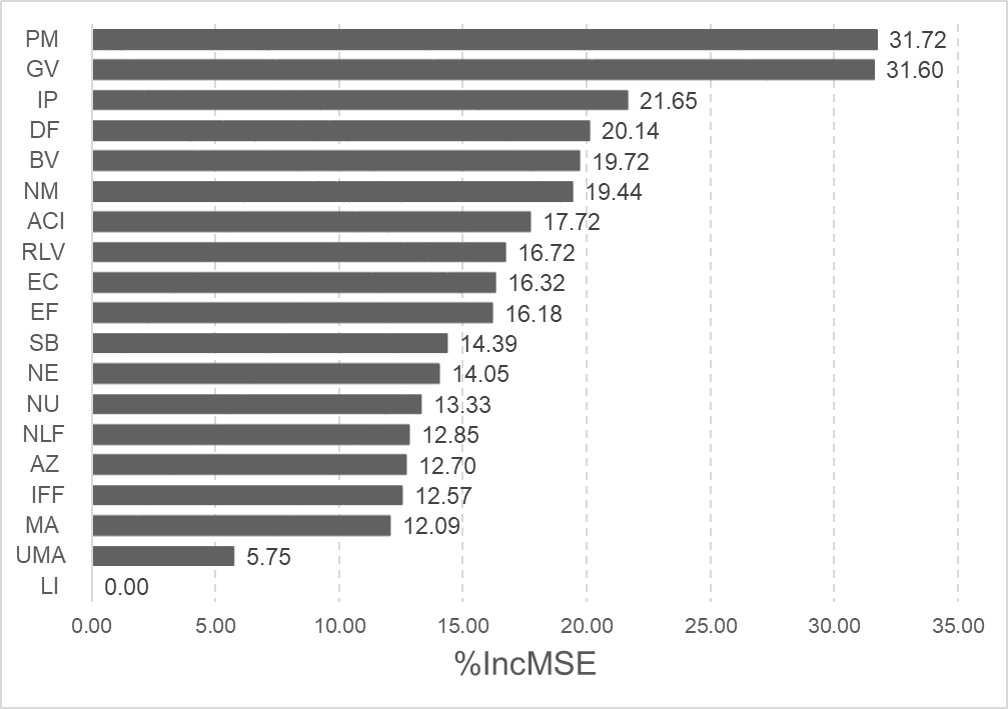}
    \caption{Importance of variables used in random forest modeling.}
    \label{fig:variableImportance}
\end{figure}

\major{As reported in Section~\ref{sec:analysisMethod}, we used Random Forest regression to predict gas consumption based on the values of the GasMet metrics. This analysis allowed us to estimate the importance of the different metrics in predicting gas consumption. Specifically, the Random Forest regression algorithm achieved an average MAE of 0.0042, an average RMSE of 0.0096, and an average R$^2$ of 0.71. This means that, on average, the trained models can explain more than 70\% of the variance in the gas consumption. The analysis of variable importance (see Figure~\ref{fig:variableImportance}) shows that all the metrics except UMA and LI exhibit an increase in Mean Square Error (MSE) higher than 12\%. More specifically, the most important variables influencing the gas consumption are the number of public members (PM) and the number of global variables (GV), both increasing the MSE by more than 30\% (31.72\% and 31.60\%, respectively) when randomly permuted. It is worth noticing that PM and GV are also the metrics exhibiting the highest correlation coefficient values with gas consumption. This result confirms that the numbers of public members (PM) and global variables (GV) might both have a significant impact on the gas consumption predictions. Thus, they are the most important metrics developers should monitor. In addition, we also report an increase of MSE higher than 20\% when the the number of indexed parameters (IP) or the number of defined functions (DF) are randomly permuted.
}

\begin{center}
\fcolorbox{black}{lightgray}{
\begin{minipage}[center]{0.97\linewidth}
\textbf{RQ$_2$ Summary:} \em{Thirteen metrics of the GasMet suite exhibit large \revise{(PM, GV, IP, NE)} or medium \revise{(ACI, AZ, IFF, RLV, NLF, NU, MA, EC, BV)} correlations with gas consumption required by smart contract deployment. 
The correlations between the pairs of GasMet metrics allow identifying frequent coding patterns that influence gas consumption.
\major{PM and GV are the most important metrics to monitor, as they might both affect gas consumption estimations significantly.
}

} 
\end{minipage}}
\end{center}

\section{Threats to validity}
\label{sec:threats}
\textit{Threats to construct validity} concern the relationship between theory and observation. The most important threat that could affect our results is related to possible imprecision/incompleteness in identifying \textit{cost smells}. In particular, such smells have been identified  by relying on information encompassed in specialized forums/books focused on the development of Solidity smart contracts. Thus, some of the identified smells (and, consequently, the related metrics in our suite) could be related to anecdotal observations. To partially mitigate this weakness, in our RQ$_1$, we studied if domain experts (\textit{i.e.,} smart contract developers) perceive such smells as relevant. The majority of survey respondents generally agree with the identified smells. However, we asked developers to rate the importance of smells by only providing short descriptions of the problems. To counteract this issue, such descriptions were accompanied by some explanatory examples. Indeed, no respondents indicated possible misunderstanding in the questions. 

\textit{Threats to conclusion validity} concern the relationship between treatment and outcome. Appropriate, non-parametric statistical procedures have been adopted to draw our conclusions concerning RQ$_2$. 
More specifically, we used the Spearman rank correlation coefficient, to investigate the relationships between the different metrics in the GasMet suite and the gas cost. To cope with multiple tests, Holm's correction procedure has been adopted to adjust p-values.

\textit{Threats to internal validity} concern factors that can affect our results. \major{The smart contract dataset considered in our study comprises small to medium size smart contracts (see Table \ref{tab:SLOC}), and this could have reduced the likelihood of specific cost smells being present in a given smart contract. In particular, such an issue may have hindered the statistical relevance of some tests. 
Indeed, there are three metrics (\ie UMA, SB, and NM) for which we only found a \textit{small} effect size in the correlation with gas consumption ($\rho < 0.3$), while our results show that for two metrics (\ie DF and EF) the correlation with gas consumption was not statistically significant ($p > 0.05$). However, the results of the analysis for estimating the importance of the GasMet metrics when used to predict gas consumption (see Figure \ref{fig:variableImportance}) showed that all the aforementioned metrics (with the exception of UMA), if randomly permuted, might have a significant impact on the gas consumption predictions (\ie $\%IncMSE > 12\%$).
To cope with this problem, in the future, we plan to replicate our study at a larger scale by also considering smart contracts of larger size.} 

\textit{Threats to external validity} concern the generalization of the findings. The set of identified cost smells is surely incomplete. Further research is needed to more-in-depth explore broader sets of Solidity coding practices that can negatively influence gas consumption. Indeed, our work jointly attempts to (i) tackle the problem of cost smells in Solidity source code, and (ii) conceive approaches to help developers more easily identifying them during smart contract development. 
With regards to RQ$_2$, our study has been carried out on a data collection comprising 2,186 real-world Solidity smart contracts for which Etherscan provides the source code. The smart contracts in our dataset may be not representative of all smart contracts deployed on the blockchain, and some of the findings may depend on the specific data we used. \change{For partially alleviating this threat, we collected a dataset that is (i) a statistically significant sample of the smart contracts for which Etherscan provides the source code, and (ii) sufficiently large to be representative of the smart contracts actually deployed on Ethereum.} However, while this study is only observational, in the future we plan to carry out experiments at a larger scale to verify the generalizability of the obtained results.
The Ethereum platform and Solidity are constantly evolving at a fast pace~\cite{DBLP:conf/wcre/WohrerZ18} and future optimizations might be applied in opcodes and/or in the compilation process of Solidity smart contracts. Clearly, the latter could have effects on the relationships exhibited by some of the metrics in our suite and the gas consumption. 
Furthermore, as our research is not exhaustive, in the future, additional metrics better outlining the code quality of smart contracts (from the gas consumption perspective) may be identified.

\section{Conclusion}
\label{sec:conclusions}
Although Blockchain technology has been established as the enabling layer for allowing the transactions of electronic cash, namely cryptocurrency, without the brokerage of a financial institution, it is now increasingly applied to many other domains. One of the key aspects to govern when developing a distributed application (dApp), i.e. an application built on the top of a DLT, is the cost of execution that, if not properly limited, can easily lead to relevant diseconomy, especially considering the issues related to guaranteeing the service levels when scaling up the distributed application. The back-end logic of a dApp is defined in smart contracts that run on the blockchain. Currently, to the best of authors' knowledge, there are no tools that can be used \textit{while} coding to help developers properly identifying the code segments that need optimizations for achieving lower gas consumption.
Considering that the choices are done by the developer while writing the smart contracts can affect the deployment and execution costs, we identified 19 patterns of code that can increase (or reduce) the gas consumption, namely \textit{cost smells}. Through a survey involving real smart contract developers, we demonstrated that the majority of respondents perceive 15 out of 19 smells as relevant. The vast majority of respondents also agree or strongly agree on the usefulness of a suite of metrics for more easily identifying such cost smells.  

On top of the identified smells, we defined a set of metrics, namely the \textit{GasMet} suite, in which each metric tries to capture the occurrences of a cost smell. Through a study involving 2,186 smart contracts, we empirically demonstrate that a subset of GasMet strongly correlates  with the gas consumption, namely \textit{GV}, \textit{PM}, \textit{IP}, and \textit{NE}, while associations with \textit{medium} effect size are observed between a further subset of the defined metrics, namely \textit{ACI, AZ, IFF, RLV, NLF, NU, MA, EC}, and \textit{BV}, and the deployment cost.
Our suite can be acquired as a tool for allowing developers to optimize smart contracts by easily localizing cost smells and improving the related code segments. \change{In particular, since we found significant links between the metrics in our suite and deployment costs, the proposed metrics can be beneficial for especially novice smart contract developers~\cite{DBLP:journals/smr/AjienkaVC20}. \textit{GasMet} metrics will guide inexperienced developers on source code portions that could be modified or refactored for reducing deployment cost. Besides, since the \textit{GasMet} suite statically collects smart contract-related metrics, it can also be used by developers and project managers as a tool for evaluating the code quality of alternative solutions from the gas consumption perspective.}

This paper provides several contributions to the research community:
\begin{itemize}
    \item a catalog of cost smells for Solidity programming language, \change{whose relevance has been assessed} through a survey involving smart contract developers (see Sections \ref{sec:elicitation} and \ref{sec:studyRelevance}); 
    \item a corresponding measurement suite, namely \textit{GasMet} for easily identifying cost smells while coding (see Section~\ref{sec:suite});
    \item an empirical \change{evaluation} of the GasMet suite, along with a corpus of smart contracts that can be used for further experiments, accessible from our replication package; and
    \item a tool that computes the GasMet metrics by statically inspecting the Solidity code of smart contracts (\major{see Section~\ref{sec:suite}}).
\end{itemize}

\change{While this study aims at (i) evaluating the perceived relevance of the identified cost smells, and (ii) proposing a set of metrics for better estimating the gas required for the deployment of the smart contracts, future work will focus on empirically setting accurate thresholds for these metrics. Since establishing robust threshold values for source code metrics is not a trivial task~\cite{DBLP:conf/icst/FontanaMMST11}, further research is needed to enable the accurate detection of the proposed smells.
\major{Actually, some of the identified code smells could be fixed directly by the Solidity compiler, which could lift the burden of the modifications off of the developer. Of course, not all the changes can be automated, since in some cases, keeping a smell might be beneficial in the economy of the overall system. For this reason, in the future, we want to investigate which code smells could be solved as compiler optimization and how to perform such optimizations.}
\major{As future work, we also plan to further refine our metrics to help developers more easily applying gas consumption optimizations.}
\major{In addition, future analyses will be aimed at more closely exploring the relationships existing between the increasing number of code lines and the specific degradations that might arise in gas consumption.} Furthermore, since the cost smells defined in this work are strictly dependent on the Solidity programming language, as future work, we \major{will} investigate if more general design practices, not related to a specific programming language, could be defined to achieve savings in gas consumption.}

\section*{Acknowledgment}
We thank all the participants in our survey. We gratefully thank Michele Fredella for his valuable help during the development of the GasMet tool.

\balance
\bibliographystyle{elsarticle-num} 
\bibliography{biblio.bib}

\end{document}